\documentclass[11pt]{amsart}
\usepackage{geometry}                
\usepackage{graphicx}
\usepackage{amssymb}
\usepackage{siunitx}

\title[Prediction of steady states  by  a machine learning technique]{Prediction of steady states in a marine ecosystem model by  a machine learning technique}

\author[Mahfuz, Slawig]{Sarker Miraz Mahfuz,  Thomas Slawig, Dep. Comp. Science, Kiel Interdisciplinary Centre of Marine Science KMS, Kiel University, 24098 Kiel, Germany}

\begin{document}
      
\maketitle

\begin{abstract}
We used precomputed steady states obtained by a spin-up for a global marine ecosystem model as training data to build a
mapping from the small number of biogeochemical model parameters onto the three-dimensional converged steady annual cycle. The mapping was performed by a conditional variational autoencoder (CVAE) with mass correction. Applied for test data, we show that the prediction obtained by the CVAE already gives a reasonable good approximation of the steady states obtained by a regular spin-up. However, the predictions do not reach the same level of annual periodicity as those obtained in the original spin-up data.  
Thus, we took the predictions as initial values for a spin-up. We could show that the number of necessary iterations, corresponding to model years, to reach a prescribed stopping criterion in the spin-up could be significantly reduced compared to the use of the originally uniform, constant initial value. 
The amount of reduction depends on the applied stopping criterion, measuring the periodicity of the solution. The savings in needed iterations and, thus, computing time for the spin-up ranges from 50 to 95\%, depending on the stopping criterion for the spin-up.   We compared these results with the use of the mean of the training data as an initial value. We found that this also accelerates the spin-up, but only by a much lower factor. 
\end{abstract}

\section{Introduction}
In many applications in climate science where numerical simulation models are used, the model is run into a steady, i.e., stationary or periodic state. 
This procedure is often called \textit{spin-up}. The reason for this is that, in order to perform a transient simulation, e.g., for a climate prediction, an appropriate initial state for the whole or the part of the climate system that is simulated, for example, the ocean, has to be found. 
Since such kind of initial state usually is, in most cases, not known or can only be interpolated from sparse measurement data, a run of the model for sometimes a quite big number of years model time is performed. In this run, the external forcing data needed in the model (for example, boundary conditions at the ocean-atmosphere interface) are fixed. The model is then started from some simple, maybe constant state and run until a more realistic, and up to some given accuracy, steady state is reached. Since the climate system does not have a stationary solution, in most realistic case this is a steady, i.e., an annually periodic solution. This spin-up computation may take hundreds or even thousands years of model time. The latter is the case for simulation where ocean models are involved.

When studying the global carbon cycle including its marine part, i.e., for simulation runs involving the marine ecosystem, this is also typically the case. 
In climate research, many different ecosystem models are used. This results in the necessity to on one hand assessing the quality of different models to represent real data, and on the other hand to optimize or estimate the inherent model parameters, that are barely measurable, in order to get the best version of a given model, cf. \cite{Schartau2017}. These kind of parameter optimization or estimation is often performed comparing the output of a steady model  with climatological data. 
When performing such kind of model assessment or parameter estimation, many model evaluations for different parameter settings are necessary. Since every evaluation requires a spin-up, the computational effort is quite high.

Several ideas have been used to accelerate spin-ups for marine ecosystem models.  The global marine ecosystem is  influenced both by the ocean currents and the marine biogeochemical processes. In a simulation model, the ocean circulation fields may be computed beforehand, leading to what is called an \textit{offline} run. This idea is used in the Transport Matrix Method (TMM) introduced by \cite{Khatiwala2005}, where the effect of the ocean currents on the biogeochemical tracers, i.e., transport and diffusion, are approximated and realized in averaged matrices. This reduces the runtime dramatically and has been widely used for parameter studies and optimization in marine ecosystem models \cite{Khatiwala2007}.
Another idea also used by \cite{Khatiwala2008} and in our implementation of the TMM \cite{Piwonski2016a}, is to apply Newton's method instead of a spin-up (that can be seen as a fixed-point iteration). This also accelerates the process of finding a steady solution, but needs appropriate line search procedures to globalize  convergence.

In this paper, we apply a machine learning (ML) technique to predict the steady annual cycle of a marine ecosystem model just by prescribing its biogeochemical parameters. In the constructed ML model, these very few parameters are mapped onto a three-dimensional distribution of a biogeochemical tracer that represents the steady annual cycle of the model. We use a very simple biogeochemical model taken from \cite{Kriest2010} which is implemented in our software Metos3d \cite{Piwonski2016a}. As training data we used available converged spin-up solution for parameters generated by latin hypercube sampling \cite{McBeCo79}. These data were available from  studies with a different focus \cite{PfeSla21}. 

The idea of replacing parts of climate models by ML techniques has become more and more popular. One typical example is the replacement of classical turbulence models by ML models, cf. \cite{Drygala2024}. Even complete models are replaced now by data-driven ML models, see \cite{Rackow2024} for an example in weather prediction. 
Our approach also aims at replacing a complete simulation model by a ML model. We used a very simple ecosystem model here, and also, we still need model runs for the training of the ML model. Thus, we see this investigation 
as a test case that might be extended to more complex ecosystem models. Moreover, it may serve as an input for research also in other areas in climate science where computationally expensive spin-ups are necessary.

The structure of the paper is as follows: We start by briefly describing the model structure in both the continuous and discrete settings, including the realization of the spin-up computation. In the third section we describe the used ML technique, and in the fourth one we give details of its realization. In the fifth section, we present our results. Therein, we discuss the prediction quality of the ML output and show what happens if these predictions are used as initial values of a spin-up. We present the  reduction in runtime that is possible compared to the standard setting with constant initial values, dependent on the accuracy of the spin-up. We end the paper with some conclusions.

\section{Model used for the Generation of Training, Validation  and Test Data}

A marine ecosystem model describes the interaction between the ocean
circulation and the marine biogeochemical cycles. Thus, it basically has two parts, one for  the computation of the ocean circulation that affects the distribution of the tracers (nutrients, phyto- and zooplankton etc.) in the ecosystem, the other one that models the biogeochemical interactions therein. The biogeochemical part may have different numbers of tracers, which determines its internal complexity. In contrast,   the ocean transport affects all tracers in a similar way and has to be applied to each of them, which again leads to a computational effort that is proportional to the number of tracers.

In a continuous setting, the governing equations are partial differential equations (PDEs) of convection-diffusion-recation type. Their number depends on the number of modeled tracers. For simplicity and since we use a model with just one tracer in this paper, we formulate all equations for one variable $y=y(x,t)$, the space- and time-dependent tracer. Looking for a steady annual cycle, i.e., an annually periodic solution, we arrive at the following boundary value problem:
  \begin{align}
  \label{eqn:Modelequation}
     \partial_t y(x,t)
           + \left( D (x,t) + A(x,t) \right) y(x,t)
        &= q \left( x, t, y, \vec{p} \right),&\quad 
         x \in \Omega, t \in [0,1], \\
    \label{eqn:Boundarycondition}
      \partial_n y(x,t) &= r \left( x, t, y, \vec{p} \right),
        &\quad x \in \partial \Omega, t \in [0,1],\\
        \label{eqn:Periodicity}
y(x,0)&=y(x,1),&x \in \Omega.
  \end{align}
  Here $\Omega$ denotes the three-dimensional spatial domain, $\partial\Omega$ its boundary, and $\partial_t$  and $\partial_n$ the time and outer normal derivative, respectively. Moreover, $D$ denotes the diffusion, $A$ the advection operator, and $q$ represents the usually nonlinear biogeochemical model with paraneters summarized in the vector $\vec{p}$.
  Advection and diffusion depend on ocean circulation data, typically velocity and turbulent diffusion coefficients. Thus, they are space- and time-dependent themselves.
  We normalized the time interval of one year to $[0,1]$ and formulated the problem with general mixed boundary conditions. The function $r$ may be used to model a flux at the lower boundary, e.g., to a sediment layer. Moreover, it allows to ensure mass conservation in water-column models also in areas where the water is shallow.
   In order to have the problem well-posed, the operators and functions $D,A,q$ and $r$ are assumed to be annually periodic, too.

  The above system of equations models a one-way  coupling from the circulation to the biogeochemistry, which allows for pre-computation of the  ocean currents.  
Then, the circulation data can be precomputed by an ocean model that has been run into a steady state as well.
  
  \subsection{Discrete Setting}
  For discretization of the ocean circulation, usually nonlinear  schemes are applied at least on the advection part (e.g., flux limiters). The diffusion operator $D$ is split into a horizontal part $D^h$ treated explicitly in time, and the vertical part $D^v$ treated implicitly. In order to reduce the amount of necessary data of pre-computed velocity fields and turbulent diffusion coefficients, \cite{Khatiwala2005} introduced the 
   \emph{transport matrix method} (TMM). The idea is to construct and use matrices that represent the application of diffusion and transport on the tracer(s), instead of using the velocity and turbulent diffusion data directly.
   The transport matrices are averaged over one month and separated into explicit  (for advection and horizontal diffusion) and implicit ones (for vertical diffusion). As result,  the simulation of the tracer
  transport is reduced to matrix-vector multiplications.
  
  Denoting be $j$ the discrete time step on a equidistant time-grid with step-size $\Delta t$ and using the explicit and implicit Euler method as described above, one step of the TMM, applied to the discrete tracer vector $\vec{y}_{j} \approx \left( y \left( t_{j}, x_{k} \right)\right)_{k=1}^{n_x}$, reads
\begin{align}
    \label{eqn:TMM}
    \vec{y}_{j+1} &= \mathbf{T}_{j}^{\text{imp}}
                     \left( \mathbf{T}_{j}^{\text{exp}} \vec{y}_j
                        + \Delta t \vec{q}_j \left( \vec{y}_j, \vec{u} \right)
                        \right)
                 =: \varphi_j \left( \vec{y}_j, \vec{p} \right),
        \quad j = 0, \ldots, n_t - 1.
  \end{align}
   Here, the 
  transport matrices have the form
   \begin{align}
    \label{eqn:tmexp}
    \mathbf{T}_{j}^{\text{exp}} &:= \mathbf{I} + \Delta t \mathbf{A}_j
                                    + \Delta t \mathbf{D}_j^h \in \mathbb{R}^{n_x \times n_x}, \\
   \label{eqn:tmimp}
    \mathbf{T}_{j}^{\text{imp}} &:= \left( \mathbf{I} - \Delta t \mathbf{D}_j^v
                                    \right)^{-1} \in \mathbb{R}^{n_x \times n_x}.
  \end{align}
  with  the identity matrix $\mathbf{I} \in \mathbb{R}^{n_x \times n_x}$ 
  and the spatially discretized counterparts $\mathbf{A}_j, \mathbf{D}_j^h$ and
  $\mathbf{D}_j^v$ of the operators $A, D_h$ and $D_v$ at time step $j$.
  The  monthly averaged matrices
  are interpolated  linearly for each time step $j$.
  We applied transport matrices  with a global  latitudinal and longitudinal resolution of $2.8125^\circ$ and 15 vertical layers, cf. \cite{Piwonski2016a}.

\subsection{Biogeochemical Model}
\label{sec:BiogeochemicalModels}

  The biogeochemical model used in this paper is the simplest one presented in \cite{Kriest2010}, called  N model. It contains only phosphate ($\textrm{PO}_4$) as inorganic nutrients. The available nutrients and light
  restrict the phytoplankton production (or biological uptake) according to the formula  \begin{align}
    \label{eqn:Phytoplankton}
    f_P: \Omega \times [0,1] \rightarrow \mathbb{R},
      f_P (x, t) &= \mu_P y_P^* \frac{I(x,t)}{K_I + I(x,t)}
                    \frac{y (x,t)}{K_N + y (x,t)}
  \end{align}
  with a maximum production rate $\mu_P$ and  an implicitly
  prescribed concentration of phytoplankton $y_P^* =
  0.0028$~  \unit{mmol\, P\, m^{-3}}. Altogether, $n_p = 5$ model parameters
  listed in Table \ref{table:ParameterValues-Modelhierarchy} control the
  biogeochemical processes of the nutrient tracer $y$.
  \begin{table*}[tb]
    \centering
    \begin{tabular}{l l l r r r}
      \hline
      Parameter & Description & Unit& \text{Value}  & Lower  & Upper bound \\
      \hline
      $p_1=k_w$      & Attenuation  of water &   \unit{m^{-1}}   & 0.02    & 0.01  & 0.05 \\
      $p_2=\mu_P$     & Maximum growth rate &   \unit{d^{-1}}  & 2.0   &  1.0   & 4.0 \\
      $p_3=K_N$      & Half saturation    \unit{PO_4} uptake &   \unit{mmol\, P\, m^{-3}}    & 0.5   & 0.25  & 1.0 \\
      $p_4=K_I$  &   Light intensity compensation &   \unit{W\, m^{-2}}    & 30.0  &  15.0  & 60.0 \\
      $p_5=b$    & Implicit  sinking speed &   \unit{1}        & 0.858 &  0.7   & 1.5 \\
         \hline
    \end{tabular}
      \caption{Parameters  of the biogeochemical models taken from
             \cite{Kriest2010} as well as lower  and upper
              bounds for their values used to generate the
             Latin hypercube samples for training, validation and test data sets.}
    \label{table:ParameterValues-Modelhierarchy}
  \end{table*}

\subsection{Computation of steady annual Cycles}
\label{sec:SteadyAnnualCycles}

  Steady annual cycles of the model for given parameter vector $\vec{p}$ 
 are computed by a
  spin-up or pseudo-time stepping scheme
   \begin{align}
    \label{eqn:Spin-upIteration}
    \vec{y}^{\ell + 1} &= \Phi \left( \vec{y}^{\ell}, \vec{p} \right),
                          \quad \ell = 0, 1, \ldots,
  \end{align}
  where
   \begin{align}
    \Phi(\vec{y},\vec{p}) &:= \varphi_{n_t -1} \circ \ldots \circ \varphi_{0}(\vec{y},\vec{p})
  \end{align}
  summarizes the steps that represent one model year model
  and  $y^{0} \in \mathbb{R}^{n_x}$ denotes the initial state.
  
  A measure for the convergence of the spin-up is the difference between two consecutive
  iterates using the \emph{spin-up norm}
  \begin{align}
    \label{eqn:StoppingCriterion}
    \varepsilon_{\ell} := \| \vec{y}^{\ell} -\vec{y}^{\ell - 1} \|_2
  \end{align}
  for iteration (model year) $\ell \in \mathbb{N}_0$. We use the Euclidean norm here, and also later when computing differences of converged solutions.
 Use of the $L^2$ norm, taking into account the different box volumes of the spatial grid, gives different numbers, but does not change the relations of differences between  solutions. Moreover, we only computed differences  in the first time step of the year. Due to the deterministic structure of the model, the  tracer distribution in one  single time-step determines  those in the whole model year. Thus, computing differences in space-time norms taken over all time-steps of the year also leads to different numbers, but to no significant change of the relations in the differences between two solutions.

\section{Machine learning technique}
We used the so-called conditional variational autoencoder (CVAE) architecture to predict steady states of the model just by the five biogeochemical model parameters. A CVAE is a generative model that allows to generate data (in our case approximations of steady states of the ecosystem model) that depend on, or are conditioned by, some other data, in our case the biogeochemical model parameters. 

An autoencoder (AE) is a pair of networks where one, the encoder, is trained to find a lower dimensional representation of the inputs, whereas the second one, the decoder, is trained to reconstruct the inputs by  this representation.
The found representation (or \textit{code} or \textit{latent variables}) may be used to compress data. The loss function that is minimized in the training is some distance function, e.g., the mean squared error, between the inputs and the outputs produced by the decoder. This basic setting is not useful for our purpose, since the latent variables have no interpretable meaning. Thus, there will be no or least no obvious and usable relation to the biogeochemical model parameters. 
In fact, not the latent variables are used to represent the biogeochemical model parameters, these  are later combined with the latent variables in order to generate different output for different biogeochemical parameters.

An AE is not a generative model, but it can be modified by considering the inputs and the latent variables as random variables or samples of them. 
This is the idea of the variational autoencoder (VAE) which is a generative model and the basis of the CVAE. Thus, we will explain the VAE concept in the following subsection before discussing the CVAE, which is then  a rather simple extension.
 We may consider  a VAE if we would want to generate steady states for one fixed parameter only.

\subsection{Variational autoencoder}
In this subsection we describe the basic idea and architecture of a VAE as well as its loss function used in the training. For more details, especially on the derivation of the loss function, we refer the reader to \cite{Kingma2022} or  \cite{Bishop2024}, from which we take the notation.

A  variational autoencoder is used as a generative model. It consists of two networks that can also be interpreted as encoder and decoder.  After training and used as a generative model, only the decoder network is applied. It produces new data based on a  random input $\vec{z}\sim\mathcal{N}(0,I_m)$, the hidden or \textit{latent} variables. Typically, they have a much smaller dimension $m$ than the data that are generated. For each  input random sample, a different output is generated. The idea is to generate outputs that are in some sense similar to given training data, but not identical to them. The decoder network, having parameters $\vec{w}$, is trained to maximize the likelihood 
\begin{align*}
p(\vec{x},\vec{w})&=\int p(\vec{x}|\vec{z},\vec{w})p(\vec{z})\,d\vec{z},
\end{align*}
where $\vec{x}$ denotes the data and $\vec{z}$ the vector of random latent variables.
This integral cannot be evaluated analytically and numerically only with high computational effort. However, for any distribution $q(\vec{z})$, the log likelihood can   be written as 
 \begin{align}
 \label{loglikeli}
\log p(\vec{x},\vec{w})&=\mathcal{L}(\vec{w})
+KL(q(\vec{z})||p(\vec{z}|\vec{x},\vec{w}))
\end{align}
 using the \textit{Kullback-Leibler divergence}
\begin{align*}
KL(q(\vec{z})||p(\vec{z}|\vec{x},\vec{w}))
&:=
-\int q(\vec{z})\log\frac{p(\vec{z}|\vec{x},\vec{w})}{q(\vec{z})}\,d\vec{z},
\end{align*}
and the \textit{evidence lower bound (ELBO)}
 \begin{align}
 \label{elbo}
\mathcal{L}(\vec{w})
&:=
\int q(\vec{z})\log\frac{p(\vec{x}|\vec{z},\vec{w})p(\vec{z})}{q(\vec{z})}\,d\vec{z}
=
\int q(\vec{z})\log p(\vec{x}|\vec{z},\vec{w})\,d\vec{z}
-KL(q(\vec{z})||p(\vec{z})).
\end{align}
   Here, the last equality is obtained by  rearranging  the terms and using the definition of the KL divergence. It can be shown that the  KL divergence is always non-negative. Thus, the ELBO in fact is a lower bound for the log likelihood.
  
The distribution $q$ can  be chosen  maximizing the ELBO and  minimizing the KL divergence. 
The idea of the VAE  is 
to train a second network, the encoder with parameters $\vec{\phi}$,
to generate an approximation whose distribution
\begin{align*}
q(\vec{z},\vec{\phi})&=\int q(\vec{z}|\vec{x},\vec{\phi})p(\vec{x})\,d\vec{x}
\end{align*}
 minimizes the KL term in the log likelihood \eqref{loglikeli}. 
 
 The actual training is  performed using the rightmost representation in \eqref{elbo}. To obtain an analytic form of the Kl divergence term, a multi-variate normal distribution for  $q(\vec{z}|\vec{x},\vec{\phi})$ is assumed or prescribed. The latter is realized by starting from $\vec{\epsilon}\sim  \mathcal{N}(0,I_m)$. The encoder network is designed to produce some $\vec{\mu},\vec{\sigma}\in\mathbb{R}^{m}$ that are interpreted as parameters of a normal distribution. By the mapping 
\begin{align}
\label{latent}
z_j&:= \mu_j+\sigma_j\epsilon_j,\quad j=1,\ldots,m,
\end{align}
 a random vector  $\vec{z}\sim \mathcal{N}(\mu,\Sigma)$ is produced,
where $\Sigma={\rm diag}\left(\sigma_j^2\right)_{j=1}^m$ is a diagonal matrix.

 As a consequence,  the KL divergence in the second representation of the ELBO in \eqref{elbo} can be evaluated exactly. The first term is approximated using a Monto Carlo sampling, and in practice with just one sample per input training data $\vec{x}_i,i=1,\ldots,n$. This results in the two terms of the loss function used in the VAE training:
\begin{align}
\label{loss-vae}
Loss_{VAE}(\vec{\phi},\vec{w})
&=
\frac1n\sum_{i=1}^n \|\vec{\hat x}_i-\vec{x}_i\|_2^2
-\frac12\sum_{j=1}^m\left(1+\log(\sigma_j^2)-\mu_j^2-\sigma_j^2\right).
\end{align}
The first term measures the quality of the reconstructions $\vec{\hat x}_i$, generated from the latent variables $\vec{z}_i$ by the decoder with parameters $\vec{w}$. The second term measures the distance between the prior distribution $p(\vec{z})$ and $q(\vec{z},\phi)$, the  output of the encoder network where $\vec{\mu}$ and $\vec{\sigma}$ depend on $\vec{\phi}$.

\begin{figure}[t]
\includegraphics[width=8.3cm]{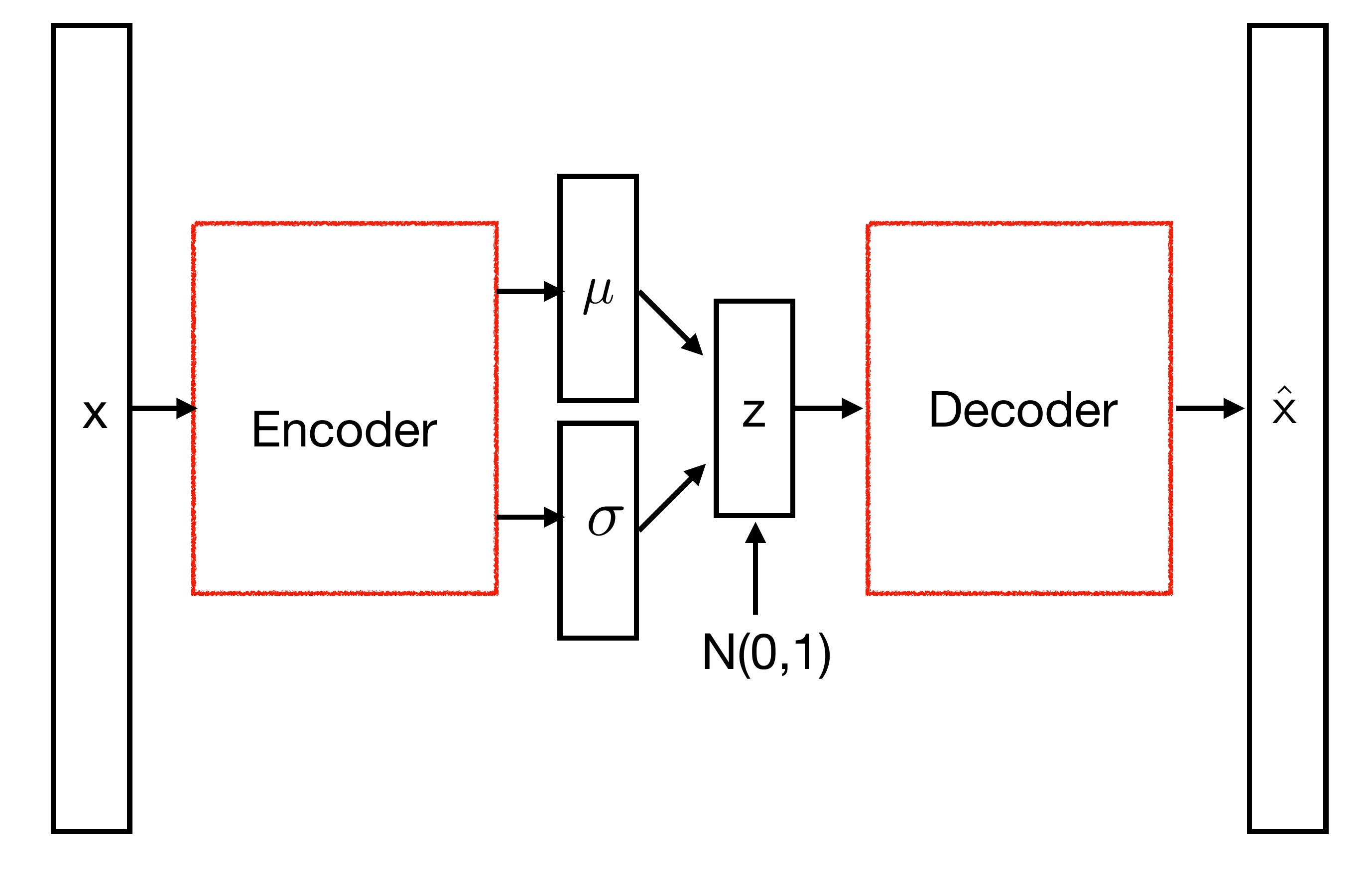}
\caption{\label{fig:vae} Basic structure of a variational autoencoder.}
\end{figure}
 
\subsection{Conditional variational autoencoder}
\label{sec:cvae}
\begin{figure}[t]
\includegraphics[width=8.3cm]{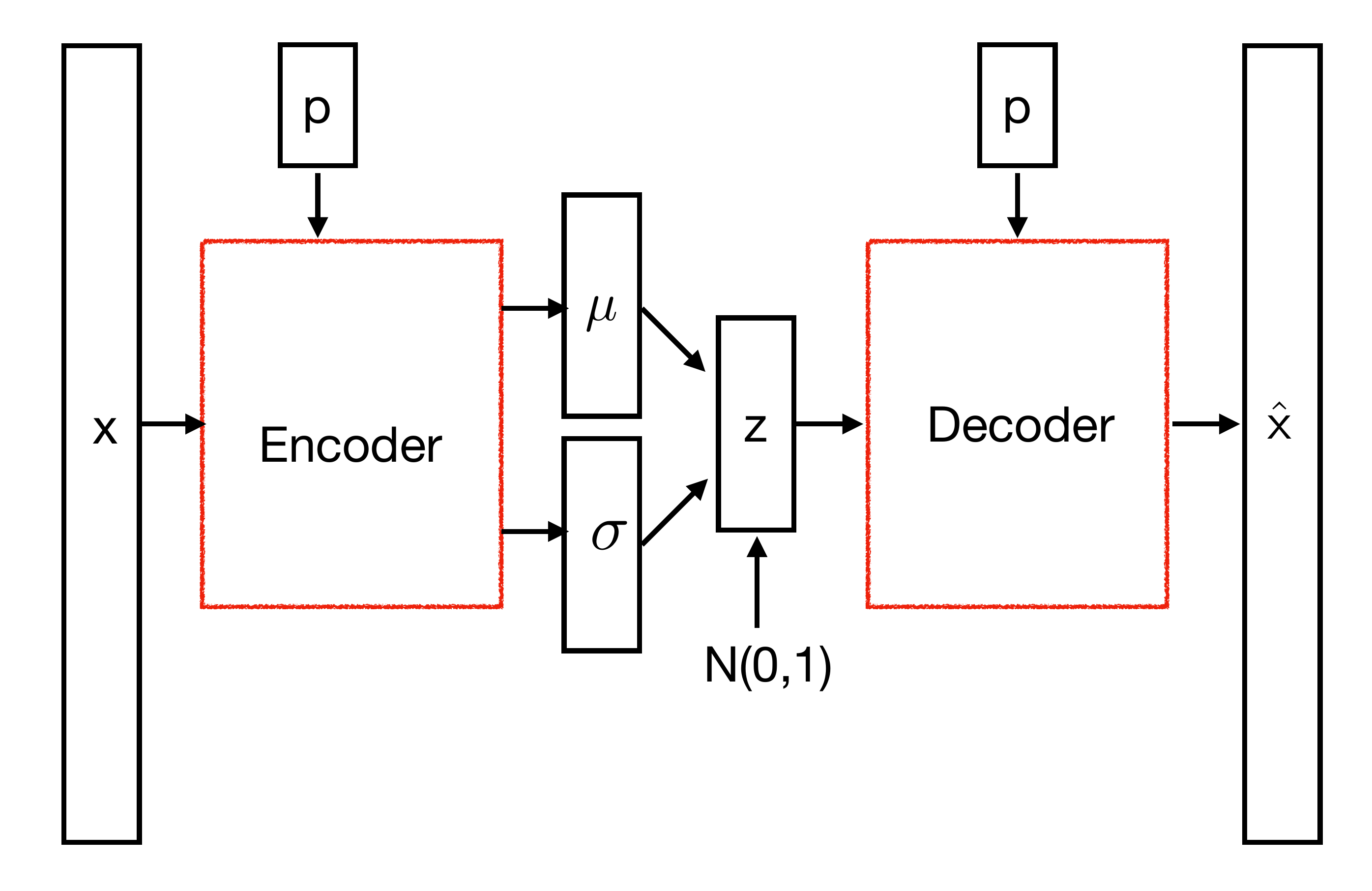}
\caption{\label{fig:cvae} Basic structure of an conditional 
variational autoencoder.}
\end{figure}

A conditional variational autoencoder (CVAE) allows  adding a label to the data that the encoder takes as input. In the same way, the decoder takes not just the  latent variables in $\vec{z}$ as input, but also the  label. Taking the label as the parameter vector of the marine ecosystem model, we can now generate a corresponding output  using the decoder. The labels, i.e., the biogeochemical parameters in our case, are just combined with the input data $\vec{x}$ for the encoder and with the latent variables $\vec{z}$ for the decoder.
The CVAE setting is shown schematically in Figure \ref{fig:cvae}. The training is done in the same way as for the VAE.
 

\section{Experimental setup}
We used 1000 samples, each  consisting of  a  parameter vector $p=\left(p_i\right)_{i=1}^{n_p},n_p=5,$ generated using Latin Hypercube Sampling \cite{McBeCo79}, and the corresponding computed  state $x\in\mathcal{X}:=\mathbb{R}^{n_\phi\times n_\lambda\times n_r},n_\phi=128,n_\lambda=64,n_r=15$.  The latter represents one time instant of an approximately steady annual cycle computed by a spin-up. Here $\lambda,\phi$ and $r$ refer to  latitude, longitude and vertical coordinate, respectively. All spin-ups for the data were run over 10'000 years of model time. From the 1000 samples, we used 600 for training, 300 for validation in training, and the last 100 for testing of the obtained CVAE model.

As mentioned in Section \ref{sec:cvae}, the main two ingredients of the used network structure are an encoder and a decoder network. 

\subsection{Encoder}
The encoder first uses a number of 3-dimensional convolutional layers, acting on the steady states $x\in\mathcal{X}=\mathbb{R}^{128\times 64\times 15}$. The input size for such a 3-d convolution layer has the size
\begin{quote}
(size 1st data dimension, size 2nd data dimension, size 3rd data dimension, number of features).
\end{quote}
The data dimensions are $(128,  64, 15)$ in our case. Parameters of a convolutional layer are the kernel size, where we used $3\times 3 \times 3$, and the strides, where  2 was taken in all three coordinate directions. For boundary points the values which are outside the 3-D data were taken to be the same as the boundary points themselves. This method is referred to as padding of type "same" in Tensorflow. The stride of 2 and this kind of padding leads to a halving of the sizes in the data dimensions by each layer (for an even value of the size).
Moreover, we used the ReLU (rectified linear unit) as activation function. Finally, the number of different filters applied in this way is specified, giving the number of features for the following layer. In the training, the convolutions are applied to a batch of several data simultaneously. In our setting, this batch size was set to 100.

In our final network design, we applied four convolutional layers. The whole encoder consists of the following steps:
\begin{itemize}
\item In the first convolutional layer, we considered the data as the only feature, setting the last input to 1, and chose the number of filters to 16. This gives an input size of $(128,64,15,1)$ which is mapped to an output of size $(64,32,8,16)$, becoming the input of the next layer.
\item The next three convolutional layers use 32, 64, and 128 filters, respectively, mapping the input to an output of size $(32,16,4,32)$, $(16,8,2,64)$, and finally  $(8,4,1,128)$.
\item Afterwards, we flatten the output to obtain a vector of size $8\times 4\times 1\times 128=4096$, and 
concatenate it with the parameter vector $\vec{p}$, giving a vector of size $4101$.
\item Now two dense layers are used to get the two  vectors $\mu$ and $\bar\sigma$, each of which has the length of 5. Using these two vectors and samples from a standard normal distribution, samples $\vec{z}\in\mathbb{R}^5$ of an $\mathcal{N}(\mu,\Sigma)$ distribution can be generated.
\end{itemize}

\subsection{Decoder}
Basically, the decoder  has the inverse structure of the encoder network described above. 
The inverse operation of a convolution is the transposed convolution. The steps in the decoder are:
\begin{itemize}
\item The sample $\vec{z}\in\mathbb{R}^5$ from the $\mathcal{N}(\mu,\Sigma)$ distribution of the encoder output is concatenated with the parameter vector $\vec{p}\in\mathbb{R}^5$, resulting in a vector $(\vec{z},\vec{p})\in\mathbb{R}^{10}$.
\item A dense layer generates a vector of size $128\times32=4096$, which is then reshaped to a tensor of size $(8,4,1,128)$.
\item Now 4 transposed 3-d convolution layers with 128, 64, 32, and 1 filter(s), respectively, are applied. They all use a kernel size of $(3,3,3)$, a stride of 2 in all directions, the same padding and the ReLU activation function as the encoder.
\end{itemize}

\subsection{Loss function}
\begin{figure}[t]
\includegraphics[width=8.3cm]{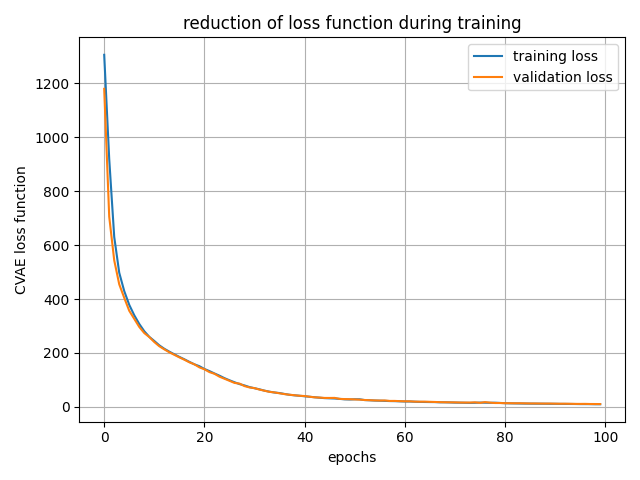}
\caption{\label{fig:training} Evolution of training and validation loss during the training process. Shown are only the first 100 of 2000 epochs.}
\end{figure}

The loss function of an CVAE used is the realization of \eqref{loss-vae} for the described data with $n=n_\phi\times n_\lambda\times n_r$.
Fig. \ref{fig:training} shows the reduction of the loss  during the training for training and validation data, which is quite uniform for both. We stopped  training after 2'000 epochs (iterations) of the ADAM optimizer of TensorFlow \cite{tensorflow2015-whitepaper}. 
 
\section{Results}
In this section, we present the results obtained for the test data by using the trained CVAE network as predictions for the steady-states. As already pointed out in Section \ref{sec:SteadyAnnualCycles}, we present results in the Euclidean norm of the differences between two spatially three-dimensional tracer distributions only. 


\subsection{Approximation quality of the predictions}
\begin{figure}[t]
\includegraphics[width=8.3cm]{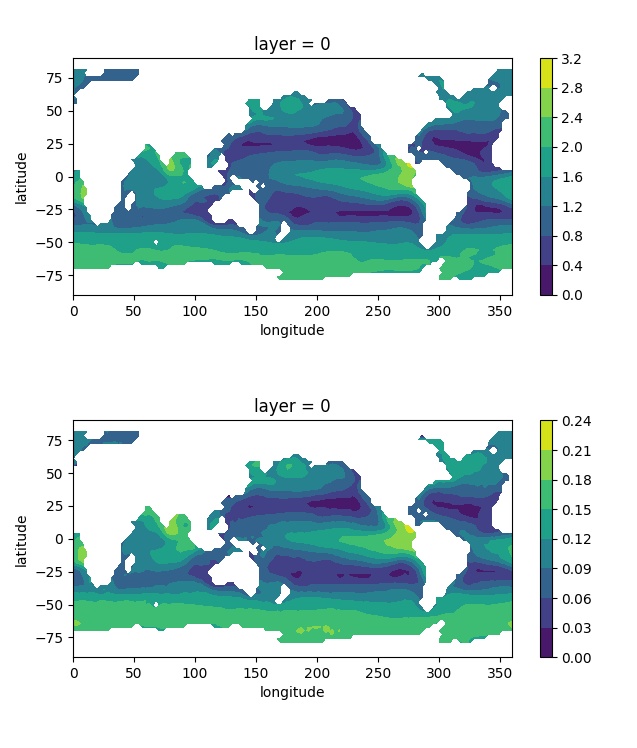}
\caption{\label{fig:results-1} Distribution of nutrients at the first time instant of the year in the uppermost layer for the original spin-up result after 10'000 years model time (top) and obtained by the prediction of the CVAE network without mass correction (bottom). Shown are results for one exemplary test parameter vector.}
\end{figure}

We start by showing the differences
of the CVAE predictions to the original converged steady states.
Fig. \ref{fig:results-1} shows one randomly chosen, exemplary 
prediction compared to the original state, both for the uppermost horizontal layer.

Due to an automatic adaption of the colors with respect to the respective maximal value, the pictures look quite similar, but the magnitude of the values is very different.
In fact, the relative difference measured in the Euclidean vector norm is about 0.92 for this example. 
 Values of same magnitude were obtained for all test data. The reason for the observed poor approximation quality, which was already observed in \cite{PfeSla22}, is a loss of total mass when applying the neural network without taking into account the mass conservation property of the model. To be more specific, the initial values of the original spin-up simulation runs are 2.17 in every grid box, resulting in a 
total mass of 2.17 when normalizing the three-dimensional volume of the ocean to 1. The differences due to numerical computation after the 10'000 years used for the spin-up are  in the range of $10^{-5}$ only for all parameters. In contrast, the mass that is contained in the CVAE predictions is about 0.17, also with only small variations among the test data.

\begin{figure}[t]
\includegraphics[width=8.3cm]{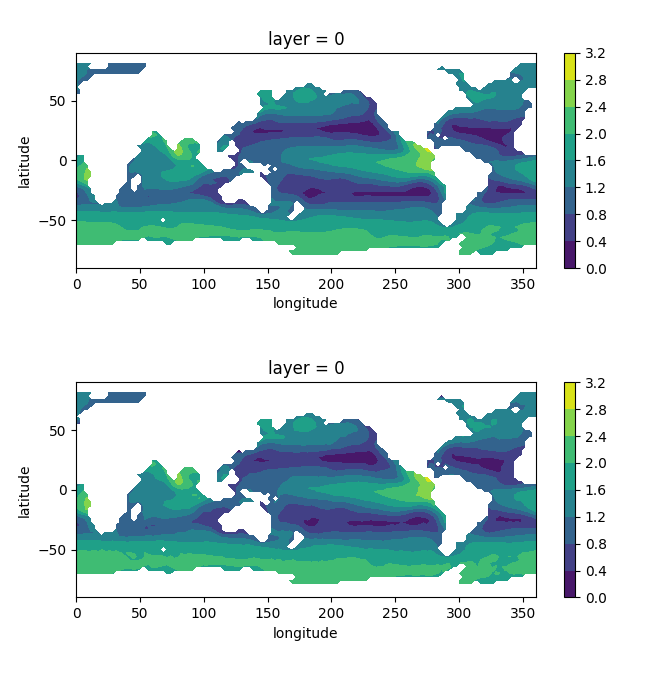}
\caption{\label{fig:results-3} Same   as  Fig. \ref{fig:results-1}, but now showing the mass-corrected prediction in the bottom picture.}
\end{figure}

\begin{figure}[t]
\includegraphics[width=8.3cm]{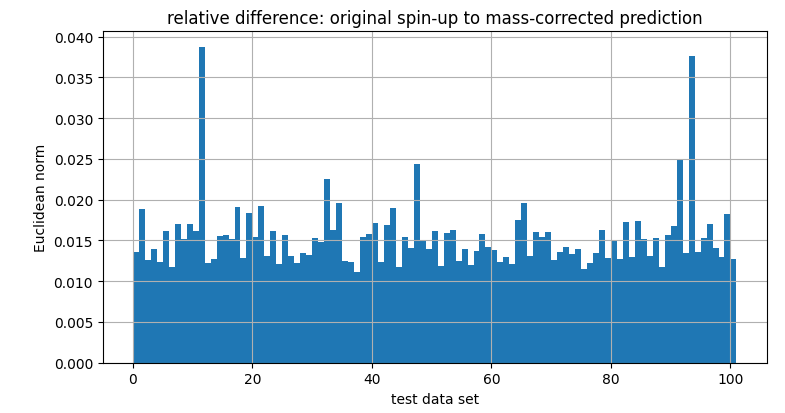}
\caption{\label{fig:results-4} Relative differences between the original spin-up results and the CVAE predictions  with mass correction.}
\end{figure}

As a remedy and analogously to the treatment in \cite{PfeSla22}, we adjusted  the mass   to the  value  2.17, the one of the initial states of the original spin-ups, by a simple post-processing step that scales the value in every box.
For the example parameter vector taken in Fig. \ref{fig:results-1}, the result can be seen  in Fig. \ref{fig:results-3}.
Now the range  of the values is close to the original spin-up result.    The relative difference  of the CVAE prediction to the result obtained by the spin-up is only about 0.013 in  Euclidean norm for this parameter vector. Fig. \ref{fig:results-4} shows that the values are similar  also for all other test data.

\begin{figure}[t]
\includegraphics[width=8.3cm]{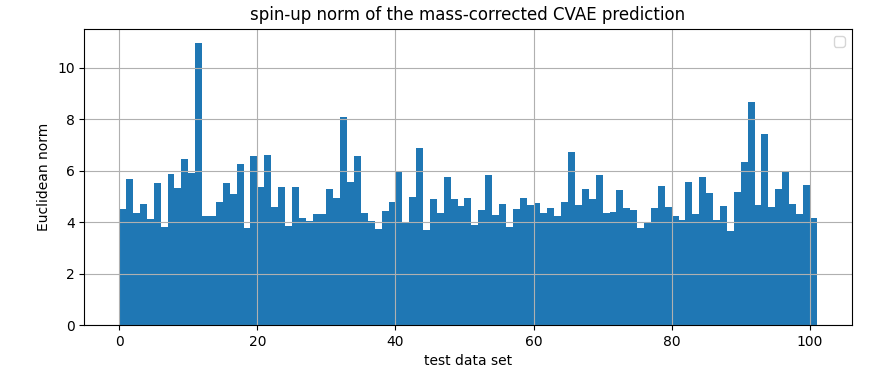}
\caption{\label{fig:results-4a} Spin-up norm of the CVAE predictions  with mass correction, obtained after computing one additional model year.}
\end{figure}

\begin{figure}[t]
\includegraphics[width=8.3cm]{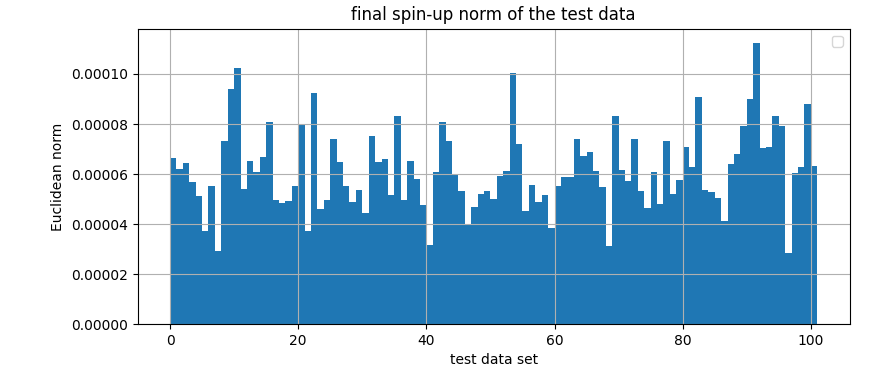}
\caption{\label{fig:results-4b} Final spin-up norm of the test data  after 10'000 model years.}
\end{figure}

The predictions seem to be rather good, however, we investigated how far they are from being steady.  
First of all, we computed one additional year and measured the difference \eqref{eqn:StoppingCriterion} afterwards. The results are show in Fig. \ref{fig:results-4a}. To interpret the values therein, we note that the final value of the spin-up norm in the test-data after a spin-up of 10'000 years are in the range of $10^{-5}$ to $10^{-4}$, which is shown in Fig. \ref{fig:results-4b}.
 
When comparing the two figures, we see that there still is a difference in the spin-up norm of four orders of magnitude, i.e., the CVAE predictions are not  
"as steady" as the original spin-up results. Thus, they may be not be used directly as a replacement of the original spin-up. However, since they are quite close, they may serve as improved initial values to reduce the necessary iterations or model years in the spin-up.

\subsection{Use of the predictions as initial values}

Since there is still quite a difference in the spin-up norm for the mass-corrected CVAE predictions, we tested how long a spin-up  takes when using these predictions as initial values. In \cite{PfeSla22}, this was also investigated using predictions from simple fully connected or sparse networks. There, the reduction in the number of iterations that was achieved and, thus, in computing time was about 13\% in average.

To estimate the possible reduction in computing time, we run  two spin-ups for each parameter vector in the test data. One of the them used a constant distribution of the tracer  as initial value,  whereas the second one starts from the mass-corrected prediction obtained by the CVAE. We stopped both spin-up runs when a given threshold or spin-up accuracy $\varepsilon$  in relative Euclidean norm was obtained. We made three series of experiments with  values of $\varepsilon =10^{-4},10^{-3}$, and $10^{-2}$, respectively. 

\begin{figure}[t]
\includegraphics[width=8.3cm]{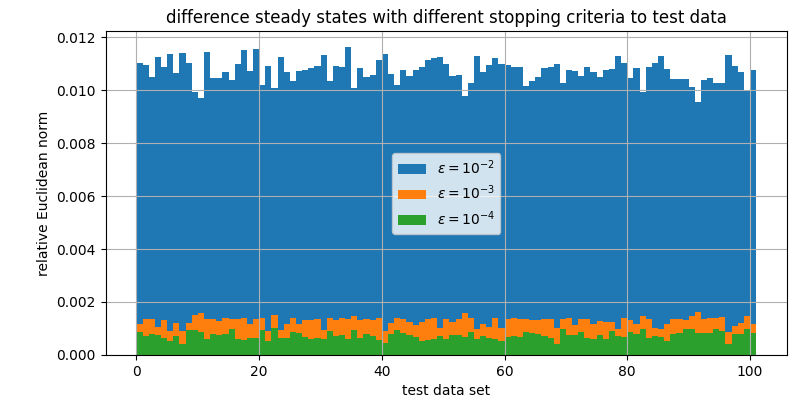}
\caption{\label{fig:results-5a} Difference of spin-up solutions with constant initial values and different stopping accuracy $\varepsilon$ compared to test data, i.e., spin-up solutions after 10'000 iterations.}
\end{figure}

To show what these three different levels of accuracy in the spin-up mean for the finally obtained, approximately steady solution,  Fig. \ref{fig:results-5a} shows
the differences of these three solutions to the original test data, which are obtained by a spin-up with 10'000 iterations. As can be already seen in Fig. \ref{fig:results-4b}, for three parameter vectors the level $\varepsilon=10^{-4}$ cannot be reached even after 10'000 iterations. From Fig. \ref{fig:results-5a}, one may deduce an appropriate stopping accuracy
which leads to a sufficient accuracy of the obtained solution. Since the differences are still rather small using $\varepsilon=10^{-3}$, this may be regarded as good compromise between accuracy and computational effort.

\begin{figure}[t]
\includegraphics[width=8.3cm]{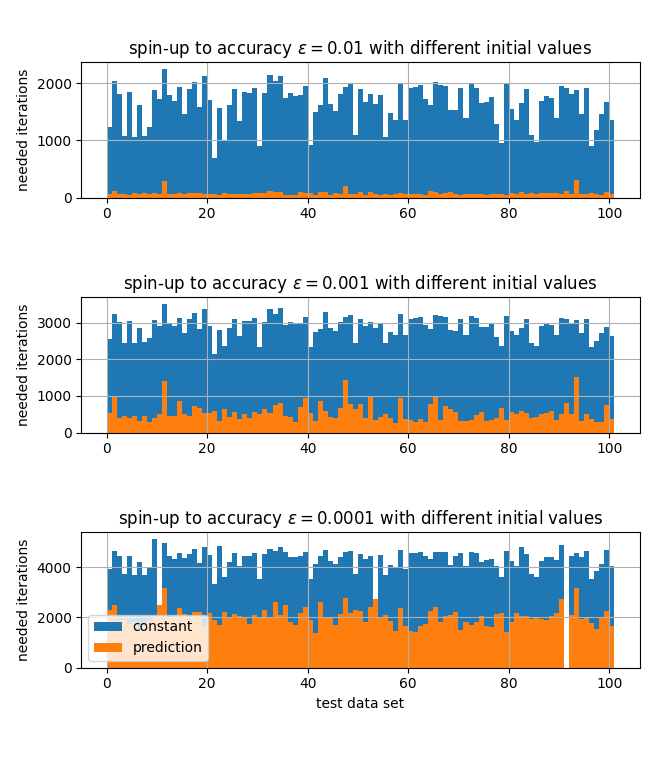}
\caption{\label{fig:results-5} Histogram showing needed iterations to reach a steady state with accuracies $\varepsilon=10^{-2},10^{-3}$  and $10^{-4}$, each for constant initial value and starting from the mass-corrected CVAE predictions. For three parameter vectors (with indices 10, 53, and 91), the spin-up does not reach the highest accuracy of $\varepsilon=10^{-4}$ for constant initial values when limiting the number of iterations to 10'000 as for the training data.  This can be seen by the missing blue bins for this test data in the bottom picture. In two of these cases, initialization with the CVAE prediction leads to convergence to this accuracy, for one parameter vector (index 91) it does not.}
\end{figure}

Now we turn to the possible savings in iterations and, thus, computing time when replacing the constant initial values by the mass-corrected CVAE predictions.
For the three stopping accuracies,  Fig. \ref{fig:results-5} shows the number of needed steps, i.e., the number  of model years needed to reach the respective accuracy in the spin-up, using the original constant initial value on one hand and the prediction by the CVAE on the other. It can be seen that the number of iterations is significantly reduced.

\begin{figure}[t]
\includegraphics[width=8.3cm]{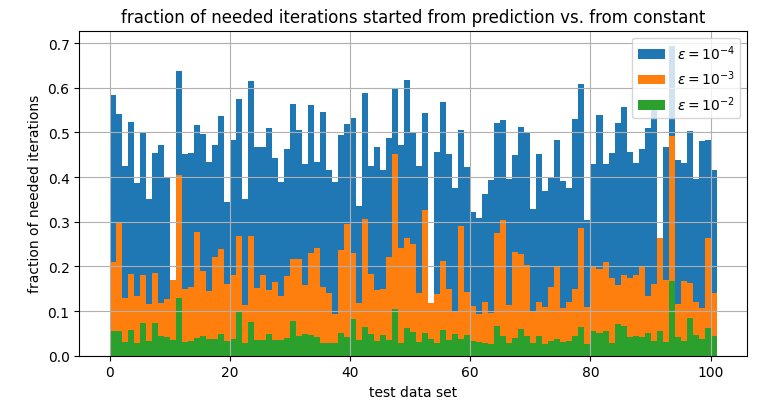}
\caption{\label{fig:results-6} Fraction of number of  needed iterations to reach a steady state when using the CVAE prediction as initial value instead of the constant one, for  different accuracies $\varepsilon$.}
\end{figure}

Fig. \ref{fig:results-6} shows the fraction of the needed steps for the three values of the accuracy when using the CVAE prediction as initial value instead of the constant one. The average values of these fractions are 0.455, 0.187 and 0.0469, i.e., there is an average reduction of over 50\%, 80\% and 95\% in the needed steps and, thus, in computing time. The saving in computing time is increasing when terminating the spin-up with lower accuracy. Reducing the spin-up accuracy naturally reduces the computation time no matter how the initial values are taken.  But using the CVAE predictions as initial values reduces the needed numbers of iterations even more when using a more relaxed stopping criterion. 

\begin{figure}[t]
\includegraphics[width=8.3cm]{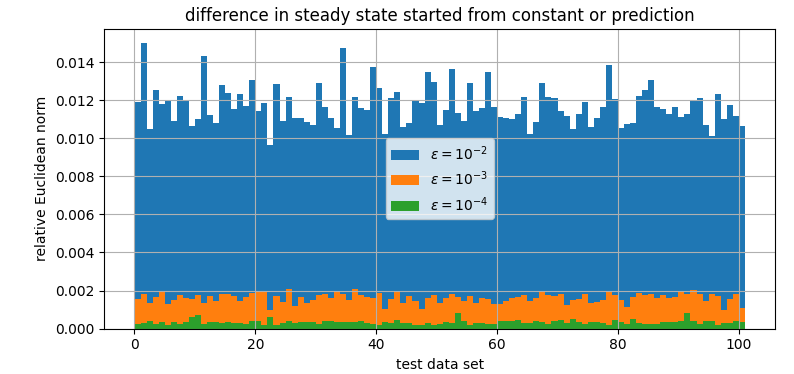}
\caption{\label{fig:results-7} Relative difference between the two converged approximately steady annual cycles (with respective accuracy $\varepsilon$), obtained with the two different initial values, a constant one and the prediction obtained by the CVAE.}
\end{figure}

To see if there are significant differences between the respective two  solutions for each stopping accuracy,  Fig. \ref{fig:results-7} depicts the difference between the converged annual cycles (with respective accuracy $\varepsilon$), obtained with the two different initial vales. 
For all three  stopping accuracies $\varepsilon$, both spin-ups give the same solution up to a reasonable difference. 
 
\begin{figure}[t]
\includegraphics[width=8.3cm]{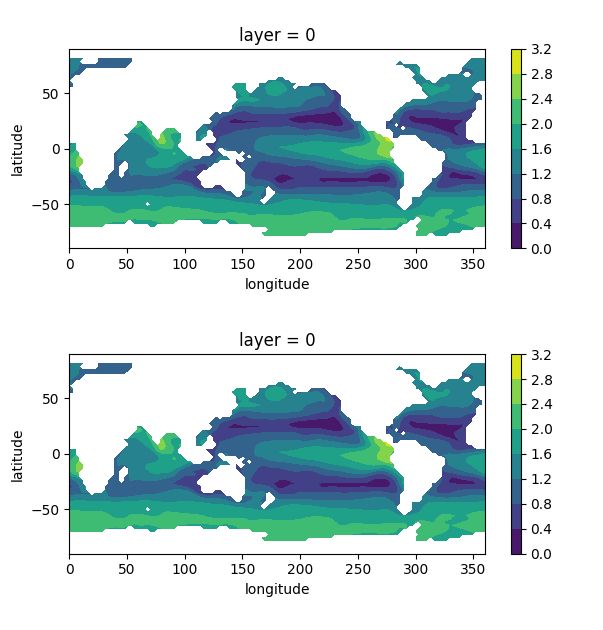}
\caption{\label{fig:results-7a} Tracer distributions, the top one obtained with constant initial value, the bottom one using the CVAE prediction, both   with stopping criterion $\varepsilon=10^{-4}$.}
\end{figure}

As an example for the accuracy of $\varepsilon= 10^{-4}$,   Fig. \ref{fig:results-7a} shows that there is no visible  difference  between the two solutions, one started from the constant initial value and the other one from the mass-corrected CVAE prediction, here for one parameter vector.
For the used model configuration, this is no surprise. In a former study \cite{PfeSla21}, it could be shown that the obtained steady state was independent of a big variety of initial value distributions, if these contain the correct overall mass.

\begin{figure}[t]
\includegraphics[width=8.3cm]{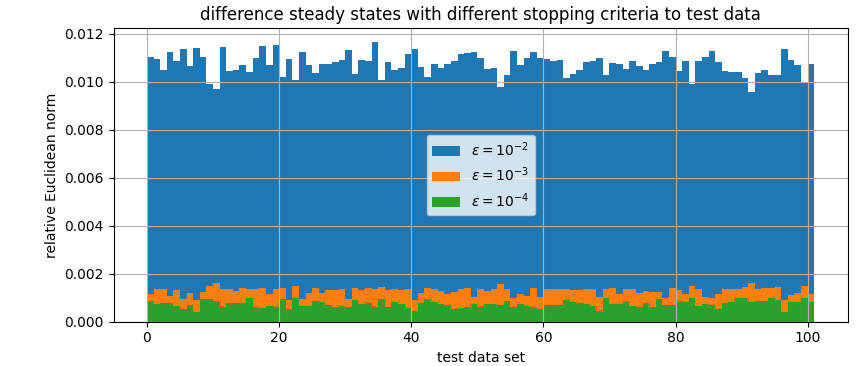}
\caption{\label{fig:results-8} Relative difference between the converged annual cycle, initialized with the CVAE prediction and stopped with an accuracy of $\varepsilon$ and the  test data, obtained with constant initial values and after 10'000 years model time.}
\end{figure}

Fig.  \ref{fig:results-8} shows that the results obtained by a spin-up, started from the mass-corrected prediction and stopped with the accuracies $\epsilon=10^{-2},10^{-3}$ and $10^{-4}$, still gives quite good approximations of the test data. 

\subsection{Using the mean of the test data as initial value}
\begin{figure}[t]
\includegraphics[width=8.3cm]{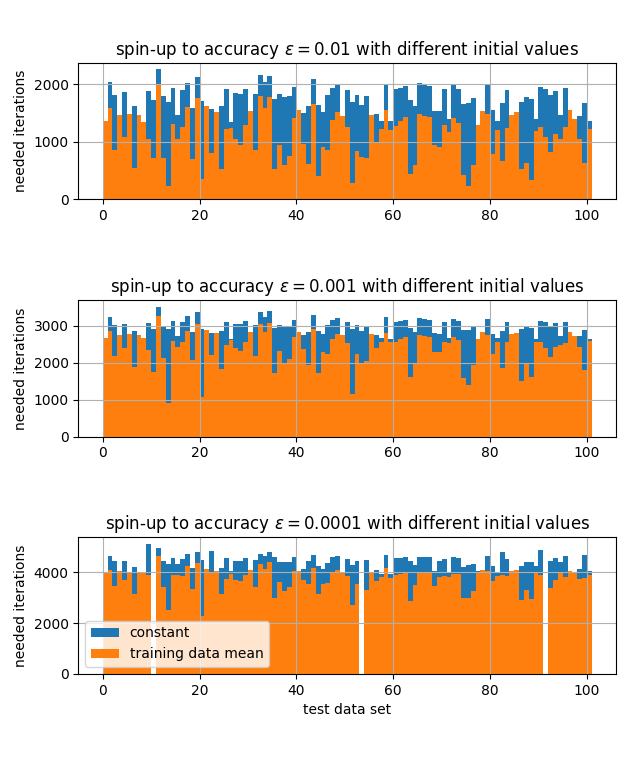}
\caption{\label{fig:results-10} Histogram showing needed iterations to reach a steady state with accuracies $\varepsilon=10^{-2}, 10^{-3}$ and $10^{-4}$, each for constant initial value and starting from the mean of the training data. For three parameter vectors (with indices 10, 53, and 91), the spin-up does not reach the highest accuracy of $\varepsilon=10^{-4}$ for both initial values when limiting the number of iterations to 10'000 as for the training data. This can be seen by the missing bins for these test data in the bottom picture.}
\end{figure}

The above results show that the spin-up can be significantly accelerated by using 
the mass-corrected predictions by the CVAE as initial values. To generate them, a training process has to be carried out. One may now argue whether there are other, maybe easier methods to extract information contained in the training data to construct a good initial value that accelerates the spin-up, too. A very simple way is to compute the mean of all training data and initialize the spin-up with this somehow artificial tracer distribution. Fig. \ref{fig:results-10} shows that this procedure accelerates the spin-up for the test data, but the effect is much smaller compared to the setting where the mass-corrected CVAE predictions are taken. The mean values of the fraction of the number of needed steps are 73, 84, and 86\%, for stopping accuracies $\varepsilon=10^{-2}, 10^{-3}$ and $10^{-4}$, respectively. Moreover, for the three parameter vectors where even after 10'000 steps the highest accuracy of $10^{-4}$ could not be reached, there is no change in this behavior, whereas when starting from the mass-corrected CVAE predictions, at least two reached this accuracy, see again Fig. \ref{fig:results-5}.
 
\section{Conclusions}
We could show that, using a machine learning technique,  it is possible to generate a model that maps the small number of biogeochemical model parameters to the three-dimensional output of the converged spin-up of a marine ecosystem model. For the studied N model, the used  conditional variational autoencoder (CVAE) with a subsequent mass correction generates three-dimensional tracer distributions  that are already close to the original results for a set of test data. Measuring the level of steadiness, i.e., the difference between solutions of two subsequent model years, it turned out that this quantity still bigger compared to those obtained by a standard spin-up procedure. However, taking the predictions as initial values for a spin-up, we could reduce the number of necessary iterations to reach a certain level of steadiness by 50 to 95\%, depending on the used stopping criterion in the spin-up. Taking the mean value of the training data as alternative initial value, we showed that this also reduces the number of needed iterations, but by a much smaller amount. 

Thus, we conclude  that the used machine learning technique may lead to a significant reduction of the computational effort for a spin-up. It might be interested to study if the number of training data can be significantly reduced, if the results also hold for more complex models with a higher number of tracers and parameters on one hand, and for fully coupled simulations that do not use  pre-computed ocean advection and diffusion operators on the other.

\subsection*{Author contribution}

Both authors developed the concept of the study. SMM tested several machine learning approaches and designs, constructed the CVAE model  and  performed the training and a preliminary evaluation. TS performed all experiments of accuracy and performance in comparison with the standard spin-up procedure and wrote the main part of the paper. SMM revised the paper and made additional contributions to the text.

\bibliographystyle{unsrt}
\bibliography{paper.bbl}
\end{document}